\documentclass[12pt]{article}
\usepackage{a4wide}
\usepackage{epsfig}
\usepackage{amsmath}

\newlength{\absize}
\catcode`@=11
\def\citer{\@ifnextchar [{\@tempswatrue\@citexr}{\@tempswafalse\@citexr[]}}

\def\@citexr[#1]#2{\if@filesw\immediate
  \write\@auxout{\string\citation{#2}}\fi
  \def\@citea{}\@cite{\@for\@citeb:=#2\do
    {\@citea\def\@citea{--\penalty\@m}\@ifundefined
       {b@\@citeb}{{\bf ?}\@warning
       {Citation `\@citeb' on page \thepage \space undefined}}%
\hbox{\csname b@\@citeb\endcsname}}}{#1}}
\catcode`@=12

\begin{document}
  \thispagestyle{empty}
  \pagestyle{empty}
  \renewcommand{\thefootnote}{\fnsymbol{footnote}}
\newpage\normalsize
    \pagestyle{plain}
    \setlength{\baselineskip}{4ex}\par
    \setcounter{footnote}{0}
    \renewcommand{\thefootnote}{\arabic{footnote}}
\newcommand{\preprint}[1]{%
  \begin{flushright}
    \setlength{\baselineskip}{3ex} #1
  \end{flushright}}
\renewcommand{\title}[1]{%
  \begin{center}
    \LARGE #1
  \end{center}\par}
\renewcommand{\author}[1]{%
  \vspace{2ex}
  {\Large
   \begin{center}
     \setlength{\baselineskip}{3ex} #1 \par
   \end{center}}}
\renewcommand{\thanks}[1]{\footnote{#1}}
\begin{flushright}
\end{flushright}
\vskip 0.5cm

\begin{center}
{\large \bf Constraint on Quantum Gravitational Well and Bose -
Einstein Statistics in Noncommutative Space}
\end{center}
\vspace{1cm}
\begin{center}
Jian-Zu Zhang$^{\ast}$
\end{center}
\vspace{1cm}
\begin{center}
Institute for Theoretical Physics, East China University of
Science and Technology, \\
Box 316, Shanghai 200237, P. R. China
\end{center}
\vspace{1cm}
\begin{abstract}
In the context of non-relativistic quantum mechanics for the case
of both position - position and momentum - momentum noncommuting,
the constraint between noncommutative parameters on the quantum
gravitational well is investigated. The related topic of
guaranteeing Bose - Einstein statistics in the general case are
elucidated: Bose - Einstein statistics is guaranteed by the
deformed Heisenberg - Weyl algebra itself, independent of
dynamics. A special feature of a dynamical system is represented
by a constraint between noncommutative parameters. The general
feature of the constraint for any system is a direct
proportionality between noncommutative parameters with a
coefficient depending on characteristic parameters of the system
under study. The constraint on the quantum gravitational well is
determined up to an arbitrary dimensionless constant.
\end{abstract}

\begin{flushleft}
${\ast}$ E-mail: jzzhang@ecust.edu.cn
\end{flushleft}

\clearpage
Physics in noncommutative space \citer{CDS,DN} has been
extensively investigated in literature. This is motivated by
studies of the low energy effective theory of D-brane with a
nonzero Neveu - Schwarz $B$ field background. Furthermore, there
is some argument that spatial noncommutativity may arise as a
quantum effect of gravity. Effects of spatial noncommutativity are
apparent only near the string scale, thus we need to work at a
level of noncommutative quantum field theory. But based on the
incomplete decoupling mechanism one expects that quantum mechanics
in noncommutative space (NCQM) may clarify some low energy
phenomenological consequences, and lead to qualitative
understanding of effects of spatial noncommutativity. In
literature NCQM  and its applications \citer{CST,JLR} have been
studied in detail. But some important questions, such as the
guarantee of Bose - Einstein statistics in the general case, have
not been resolved.

Recently the quantum gravitational well has attracted attentions
\cite{Nesv02,Nesv03,Nesv05,Bert05a}. The existence of quantum
states of particles in the gravitational field, as ones in
electromagnetic and strong fields, is  expected for a long time.
The lowest stationary quantum state of neutrons in the Earth's
gravitational field is identified in the laboratory. In the case
of noncommutative space it is noticed that, because of the
speciality of its linear potential \cite{Bert05a}, the constraint
between noncommutative parameters on the quantum gravitational
well is not clear.

In this paper the discussions are restricted in the context of
non-relativistic quantum mechanics.
In this paper our attention focuses on elucidating this topic,
which is closely related to the above question about guaranteeing
Bose - Einstein statistics in the general case. We find that Bose
- Einstein statistics is guaranteed by the deformed Heisenberg -
Weyl algebra itself, independent of dynamics. A special feature of
a dynamical system is represented by a constraint between
noncommutative parameters. The general feature of the constraint
for any system is a direct proportionality between noncommutative
parameters with a coefficient depending on characteristic
parameters of the system under study. Such a constraint for the
quantum gravitational well is fixed up to a dimensionless
constant.

{\bf The Deformed Heisenberg - Weyl Algebra} - In the following we
review the background first. In order to develop the NCQM
formulation we need to specify the phase space and the Hilbert
space on which operators act. The Hilbert space is consistently
taken to be exactly the same as the Hilbert space of the
corresponding commutative system \citer{CST}. There are different
types of noncommutative theories, for example, see a review paper
\cite{DN}.

As for the phase space we consider both position - position
noncommutativity (space-time noncommutativity is not considered)
and momentum - momentum noncommutativity. In this case the
consistent deformed Heisenberg - Weyl algebra \cite{JZZ04a} is:
\begin{equation}
\label{Eq:xp}
[\hat x_{i},\hat x_{j}]=ib\theta\epsilon_{ij},
\qquad [\hat p_{i},\hat p_{j}]=ib\eta\epsilon_{ij}, \qquad
[\hat x_{i},\hat p_{j}]=i\hbar\delta_{ij},\;(i,j=1,2),
\end{equation}
where $\theta$ and $\eta$ are constant parameters, independent of
the position and momentum.
Here we consider the intrinsic momentum - momentum
noncommutativity. It means that the parameter $\eta$, like the
parameter $\theta$, should be extremely small. This is guaranteed
by a direct proportionality provided by a constraint between them
(See below).
The $\epsilon_{ij}$ is a two-dimensional antisymmetric unit
tensor, $\epsilon_{12}=-\epsilon_{21}=1,$
$\epsilon_{11}=\epsilon_{22}=0$. In Eq.~(\ref{Eq:xp}) $b$ is a
dimensionless constant, which will be fixed later.

In literature the deformed Heisenberg - Weyl algebra (\ref{Eq:xp})
has different realizations by undeformed variables
\cite{NP,Bert05a}. Here we show that representations of deformed
variables $\hat x_{i}$ and $\hat p_{i}$ by undeformed variables
$x_{i}$ and $p_{i}$ are {\it uniquely} fixed by a requirement of
consistency of the framework. We consider the following ansatz
(Henceforth summation convention is used)
\begin{equation}
\label{Eq:hat-x-p}
\hat x_{i}=\xi(x_{i}-\frac{1}{2\hbar}\theta\epsilon_{ij}p_{j}),
\quad
\hat p_{i}=\xi(p_{i}+\frac{1}{2\hbar}\eta\epsilon_{ij}x_{j}),
\end{equation}
where $x_{i}$ and $p_{i}$ satisfy the undeformed Heisenberg - Weyl
algebra
$$[x_{i},x_{j}]=[p_{i},p_{j}]=0,\; [x_{i},p_{j}]=i\hbar\delta_{ij}.$$
In Eq.~(\ref{Eq:hat-x-p}) parameter $\xi$ is a dimensionless
constant, which can be fixed as follows. Inserting
Eqs.~(\ref{Eq:hat-x-p}) into the first equation of
Eqs.~(\ref{Eq:xp}), it follows that $[\hat x_{i},\hat
x_{j}]=ib\theta\epsilon_{ij}=i\xi^2\theta\epsilon_{ij}$, thus
$b=\xi^2$ (For the equation $[\hat p_{i},\hat
p_{j}]=ib\eta\epsilon_{ij}$, we obtain the same result).
Furthermore, the Heisenberg commutation relation $[\hat x_{i},\hat
p_{j}]=i\hbar\delta_{ij}$ should be maintained by
Eqs.~(\ref{Eq:hat-x-p}). Inserting Eqs.~(\ref{Eq:hat-x-p}) into
the third equation of Eqs.~(\ref{Eq:xp}), we obtain that $[\hat
x_{i},\hat p_{j}]=i\hbar\delta_{ij}=i\hbar
\xi^2(1+\theta\eta/4\hbar^2)\delta_{ij}$. Thus both parameters $b$
and $\xi$ are fixed:
\begin{equation}
\label{Eq:xi}
\xi=(1+\theta\eta/4\hbar^2)^{-1/2},\; b=\xi^2.
\end{equation}
The parameter $\xi$ is called the scaling factor. When $\eta=0,$
we have $\xi=1$. The deformed Heisenberg - Weyl algebra
(\ref{Eq:xp}) reduces to the one of only position - position
noncommuting.
For the case of both position - position and momentum - momentum
noncommuting the scaling factor $\xi$ plays a role for
guaranteeing consistency of the framework.

An another choice of the scaling factor leads to the relations
obtained in Ref.~\cite{Bert05a} with the Planck constant $\hbar$
replaced by an effective Planck constant
$\hbar_{eff}=\hbar(1+\theta\eta/4\hbar^2)$, which are consistent
with the same algebra (\ref{Eq:xp}).

It is worth noting \cite{JZZ05a} that, unlike the case of only
position - position noncommuting, the determinant $\mathcal{R}_s$
of the transformation matrix $R_s$ between $(\hat x_1,\hat
x_2,\hat p_1,\hat p_2)$ and $(x_1,x_2,p_1,p_2)$ in
Eqs.~(\ref{Eq:hat-x-p}) is
$\mathcal{R}_{s}=\xi^4(1-\theta\eta/4\hbar^2)^2$.
When $\theta\eta=4\hbar^2$, the matrix $R_s$ is singular. This
means that the deformed variables $(\hat x_{i},\hat p_{i})$ and
the undeformed variables $(x_{i},p_{i})$ are not completely
equivalent in physics.

{\bf The Quantum Gravitational Well} - Recently, the lowest
stationary quantum state of a particle in the Earth's
gravitational field is identified in the laboratory
\citer{Nesv02,Nesv05}. The experimental setup is as follows. A
system of a particle of mass $\mu_n$ in the potential well formed
by the two dimensional constant Earth's gravitational field, ${\bf
g}=-g{\bf e_x}$, where $g$ is the standard gravitational
acceleration at the sea level, and a horizontal mirror placed at
$x=0$.  This system is known as the quantum gravitational well. In
the laboratory the gravitational field alone does not creating a
potential well, as it can only confine particles to fall along
gravity lines. Thus a horizontal mirror is necessary for creating
the well. In order to avoid that electromagnetic effects overlap
the effect of the Earth's gravitational field, in the experiment
some neutral particles with a long lifetime, such as slow
neutrons, are chosen.

If NCQM is a realistic physics, low energy quantum phenomena
should be reformulated in terms of the deformed operators. The
Hamiltonian of the quantum gravitational well, formulated in terms
of the deformed phase space variables $\hat x_i$ and $\hat p_i$,
is
\begin{equation}
\label{Eq:Hg}
\hat H=\frac{1}{2\mu_n}\hat p_i^2+\mu_{n}g\hat x_1.
\end{equation}

This system exhibits both theoretical and experimental interests.
Investigations of low energy properties of this system may help to
explore effects of spatial noncommutativity, and hopefully shed
some new light on physical reality. In many systems, the potential
can be modelled by a harmonic oscillator through an expansion
about its minimum. The speciality of the gravitational well is
that its linear potential is not the case. Recently it is argued
\cite{Bert05a} that, because of its speciality, a direct
proportionality between noncommutative parameters in the
constrained condition explored in two dimensional harmonic
oscillator \cite{JZZ04a} may not apply to the quantum
gravitational well.

This problem is related to guarantee Bose - Einstein statistics in
the case of both position - position and momentum - momentum
noncommuting. We find that the maintenance of Bose - Einstein
statistics is determined by the deformed Heisenberg - Weyl algebra
itself, independent of dynamics. A special feature of a dynamical
system is represented by a constraint between noncommutative
parameter, which can be determined by the deformed bosonic
algebra. The general feature of such a constraint is a direct
proportionality between noncommutative parameters $\eta$ and
$\theta$, which works for any dynamical systems, including the
quantum gravitational well.

In the following we first investigate Bose - Einstein statistics
for the general case.

{\bf The Guarantee of Bose - Einstein Statistics} - In literature
the guarantee of noncommutative Bose - Einstein statistics in the
case of both position - position and momentum - momentum
noncommuting has not been resolved.

In this paragraph we demonstrate that for the general case
noncommutative Bose - Einstein Statistics is guaranteed by the
following existence theorem:

{\bf Theorem} {\it Bose - Einstein statistics for the case of both
position - position noncommutativity and momentum - momentum
noncommutativity is guaranteed by the deformed Heisenberg - Weyl
algebra itself, independent of dynamics. The deformed bosonic
algebra constitutes a complete and closed algebra.}

In the context of non-relativistic quantum mechanics the proof of
this theorem includes two aspects. The first aspect is to
construct the general representations of deformed annihilation and
creation operators
from the deformed Heisenberg - Weyl algebra itself, and to find
the complete and closed deformed bosonic algebra. The second
aspect is about the general construction of the Fock space of
identical bosons in noncommutative space in which the formalism of
the deformed bosonic symmetry of restricting the states under the
permutation of identical particles in multi - particle systems can
be developed.

On the level of quantum field theory the annihilation and creation
operators appear in the expansion of the field operator. In the
context of non-relativistic quantum mechanics the deformed
annihilation operator $\hat a_i$ can be generally represented by
$\hat x_i$ and $\hat p_i$ as
\begin{equation}
\label{Eq:hat-a}
\hat a_i=c_1(\hat x_i +ic_2\hat p_i),
\end{equation}
where the constants $c_1$ and $c_2$  can be fixed as follows. The
operators $\hat a_i$ and $\hat a_i^\dagger$ should satisfy the
bosonic commutation relations $[\hat a_1,\hat a_1^\dagger]=[\hat
a_2,\hat a_2^\dagger]=1$ (to keep the physical meaning of $\hat
a_i$ and $\hat a_i^\dagger$). From this requirement and the
deformed Heisenberg - Weyl algebra (\ref{Eq:xp}) it follows that
$c_1=\sqrt{1/2\hbar c_2}$. When the state vector space of
identical bosons is constructed by generalizing one - particle
quantum mechanics, Bose - Einstein statistics should be maintained
at the deformed level described by $\hat a_i$, thus operators
$\hat a_i$ and $\hat a_j$ should be commuting: $[\hat a_i,\hat
a_j]=0$. From this equation and the deformed Heisenberg - Weyl
algebra (\ref{Eq:xp}) it follows that
$ic_1^2\xi^2\epsilon_{ij}(\theta-c_2^2\eta)=0$. Thus {\it the
condition of guaranteeing Bose - Einstein statistics} reads
\begin{equation}
\label{Eq:c-2}
c_2=\sqrt{\theta/\eta}.
\end{equation}
The general representations of the deformed annihilation and
creation operators $\hat a_i$ and $\hat a_i^\dagger$ are
\begin{equation}
\label{Eq:aa+1}
\hat a_i=\sqrt{\frac{1}{2\hbar}\sqrt{\frac{\eta}{\theta}}}\left
(\hat x_i +i\sqrt{\frac{\theta}{\eta}}\hat p_i\right),
\hat
a_i^\dagger=\sqrt{\frac{1}{2\hbar}\sqrt{\frac{\eta}{\theta}}}\left
(\hat x_i-i\sqrt{\frac{\theta}{\eta}}\hat p_i\right),
\end{equation}
From Eqs.~(\ref{Eq:aa+1}) and (\ref{Eq:c-2}) it follows that the
deformed bosonic algebra of $\hat a_i$ and $\hat a_j^\dagger$
reads
\begin{equation}
\label{Eq:[a,a+]1}
[\hat a_i,\hat a_j^\dagger]=\delta_{ij}
+\frac{i}{\hbar}\xi^2\sqrt{\theta\eta}\;\epsilon_{ij},\;
[\hat a_i,\hat a_j]=0,\;(i,j=1,2).
\end{equation}
In Eqs.~(\ref{Eq:[a,a+]1}) the three equations $[\hat a_1,\hat
a_1^\dagger]=[\hat a_2,\hat a_2^\dagger]=1,\;[\hat a_1,\hat
a_2]=0$ are the same as the undeformed bosonic algebra in
commutative space; The equation
\begin{equation}
\label{Eq:[a,a+]2}
[\hat a_1,\hat a_2^\dagger] =\frac{i}{\hbar}\xi^2\sqrt{\theta\eta}
\end{equation}
is a new type. Eqs.~(\ref{Eq:[a,a+]1}) constitute a complete and
closed deformed bosonic algebra.

The second aspect is about the general construction of the Fock
space of identical bosons in noncommutative space. Following the
standard procedure of constructing the Fock space of many -
particle systems in commutative space, we shall take
Eqs.~(\ref{Eq:[a,a+]1}) as the defining relations for the complete
and closed deformed bosonic algebra without making further
reference to its $\hat x_i$, $\hat p_i$ representations,
generalize it to many - particle systems and complete the deformed
Bosonic symmetry. In the case of both position - position and
momentum - momentum noncommuting the special future is when $[\hat
a_{i},\hat a_{j}]=[\hat a_{i}^\dagger,\hat a_{j}^\dagger]=0$ are
satisfied, Bose - Einstein statistics is not yet quaranteed. The
season is as follows. Because the new bosonic commutation relation
(\ref{Eq:[a,a+]2}) correlates different degrees of freedom, the
number operators $\hat N_1=\hat a_1^\dagger\hat a_1$ and $\hat
N_2=\hat a_2^\dagger\hat a_2$ do not commute, $[\hat N_1, \hat
N_2]\ne 0.$ They have not common eigenstates. In this case the
construction of the Fock space is involved.
\footnote {\; The definition of the vacuum state $|0,0\rangle$ is
the same as in commutative space,
$\hat a_i|0,0\rangle=0,\;(i=1,2)$.
If a general hat state $\widehat {|m,n\rangle}$ is defined as
$\widehat {|m,n\rangle}\equiv c(\hat a_1^\dagger)^m(\hat
a_2^\dagger)^n|0,0\rangle$ where $c$ is the normalization
constant,
 these states $\widehat {|m,n\rangle}$
are not the eigenstate of $\hat N_1$ and $\hat N_2$:
\begin{equation*}
\hat N_1\widehat {|m,n\rangle} =m\widehat
{|m,n\rangle}+\frac{i}{\hbar}m\xi^2 \sqrt{\theta\eta}\widehat
{|m+1,n-1\rangle}, \nonumber
\end{equation*}
\begin{equation*}
\hat N_2\widehat {|m,n\rangle} =n\widehat
{|m,n\rangle}+\frac{i}{\hbar}n\xi^2 \sqrt{\theta\eta}\widehat
{|m-1,n+1\rangle}. \nonumber
\end{equation*}
Because of the new bosonic commutation relation
(\ref{Eq:[a,a+]2}), in calculations of the above equations we
should take care of the ordering in the state $\widehat
{|m,n\rangle}$. The states $\widehat {|m,n\rangle}$ are not
orthogonal each other. For example, the inner product between
$\widehat {|1,0\rangle}$ and $\widehat {|0,1\rangle}$ is:
\begin{equation*}
\label{Eq:1-2b} \widehat {\langle 1,0|}\widehat {
1,0\rangle}=-\frac{i}{\hbar}\xi^2 \sqrt{\theta\eta}. \nonumber
\end{equation*}
Thus $\{\widehat {|m,n\rangle}\}$ do not constitute an orthogonal
complete basis.}
A full investigations of noncommutative Bose - Einstein statistics
should complete the deformed Bosonic symmetry under the
permutation of identical particles. Thus we should successfully
construct the Fock space of identical bosons.

We introduce the following tilde annihilation and creation
operators
\begin{equation}
\label{Eq:tilde-a}
\tilde a_1=\frac{1}{\sqrt{2\alpha_1}} \left(\hat a_1+i\hat
a_2\right),\;
\tilde a_2=\frac{1}{\sqrt{2\alpha_2}} \left(\hat a_1-i\hat
a_2\right),
\end{equation}
where $\alpha_{1,2}=1\pm\frac{i}{\hbar}\xi^2\sqrt{\theta\eta}$.
From Eqs.~(\ref{Eq:[a,a+]1}) it follows that the commutation
relations of $\tilde a_i$ and $\tilde a_j^\dagger$ read
$\left[\tilde a_i,\tilde a_j^\dagger\right]=\delta_{ij},\;$
$\left[\tilde a_i,\tilde a_j\right]=\left[\tilde
a_i^\dagger,\tilde a_j^\dagger\right]=0,\;(i,j=1,2)$.
Thus $\tilde a_i$ and $\tilde a_i^\dagger$ are explained as the
deformed annihilation and creation operators in the tilde system.
The tilde number operators $\tilde N_1=\tilde a_1^\dagger\tilde
a_1$ and $\tilde N_2=\tilde a_2^\dagger\tilde a_2$ commute each
other, $[\tilde N_1,\tilde N_2]= 0.$ The state $|0,0\rangle$ is
also the vacuum state in the tilde system. A general tilde state
$\widetilde {|m,n\rangle}\equiv (m!n!)^{-1/2}(\tilde
a_1^\dagger)^m(\tilde a_2^\dagger)^n|0,0\rangle$
are the common eigenstate of $\tilde N_1$ and $\tilde N_2$:
$\tilde N_1\widetilde {|m,n\rangle}=m\widetilde {|m,n\rangle}$,
$\tilde N_2\widetilde {|m,n\rangle}=n\widetilde {|m,n\rangle}$,
$(m, n=0, 1, 2,\cdots)$.
We obtain $\widetilde {\langle m^{\prime},n^{\prime}} \widetilde
{|m,n\rangle}= \delta_{m^{\prime}m}\delta_{n^{\prime}n}$. Thus
$\{\widetilde {|m,n\rangle}\}$ constitute an orthogonal normalized
complete basis of the tilde Fock space.

In the above we proved the theorem. Now we investigate some issues
related to the theorem.

{\it The correlated bosonic commutation relation} - Different from
the case in commutative space, the new bosonic commutation
relation (\ref{Eq:[a,a+]2}) correlates different degrees of
freedom to each other, so it is called the correlated bosonic
commutation relation. It encodes effects of spatial
noncommutativity at the level of $\hat a_i$ and $\hat
a_i^\dagger$, and plays essential roles in dynamics. It is the
origin of the fractional angular momentum \cite{JZZ04a}.

The correlated bosonic commutation relation (\ref{Eq:[a,a+]2}) is
consistent with {\it all} principles of quantum mechanics and Bose
- Einstein statistics.

In literature one is not aware of the correlated bosonic
commutation relation (\ref{Eq:[a,a+]2}). This means that in
literature the deformed bosonic algebra is closed, but not
complete.

{\it Consistency of the framework} - In the above we prove that
for the case of both position - position and momentum - momentum
noncommuting, Bose - Einstein statistics is guaranteed by the
general construction of deformed annihilation and creation
operators (\ref{Eq:aa+1}). If momentum - momentum is commuting
($\eta= 0$), it is impossible to obtain $[\hat a_i,\hat a_j]=0$.
We conclude that in order to maintain Bose - Einstein statistics
for identical bosons at the deformed level we should consider both
position - position noncommutativity and momentum - momentum
noncommutativity.

{\it The Constrained Condition} - 
The structure of the deformed annihilation and creation operators
$\hat a_{i}$ and $\hat a_{i}^\dagger$ in Eqs.~(\ref{Eq:aa+1}) are
determined by the deformed Heisenberg - Weyl algebra
(\ref{Eq:xp}), independent of dynamics. The special feature of a
dynamical system is encoded in the dependence of the factor
$\sqrt{\theta/\eta}$ on characteristic parameters of the system
under study. This put a constraint on $\theta$ and $\eta$. From
Eq.~(\ref{Eq:c-2}) it follows the following constrained condition
\begin{equation}
\label{Eq:cc1}
\eta=K\theta,
\end{equation}
where the coefficient $K=c_2^{-2}$ is a constant with a dimension
$(mass/time)^2$, and depends on characteristic parameters of the
system under study. Thus the momentum - momentum noncommutative
parameter $\eta$ also depends on characteristic parameters of a
dynamical system. Such a dependency of $\eta$ on parameters of the
Hamiltonian (or the action) can be understood based on the
following observation: noncommutativity between momenta arises
naturally as a consequence of noncommutativity between
coordinates, as momenta are defined to be the partial derivatives
of the action with respect to the noncommutative coordinates
\cite{SGT}.

Eq.~(\ref{Eq:cc1}) shows that the general feature of such a
constraint is a direct proportionality between noncommutative
parameters $\eta$ and $\theta$. The deformed Heisenberg - Weyl
algebra (\ref{Eq:xp}) is foundations of noncommutative quantum
theories. A result derived from this algebra is a fundamental one.
The condition (\ref{Eq:c-2}) of guaranteeing Bose - Einstein
statistics is determined by the deformed Heisenberg - Weyl
algebra. This means that Eq.~(\ref{Eq:cc1}) is based on a
fundamental principle in noncommutative quantum theories. It is a
general result, and can apply to any dynamical systems.

{\it Realizations of the deformed annihilation and creation
operators by the undeformed ones} - The representations of the
deformed annihilation and creation operators $\hat a_i$ and $\hat
a_i^\dagger$ by the undeformed ones $a_i$ and $a_i^\dagger$ are
consistently determined by Eqs.~(\ref{Eq:xp})- (\ref{Eq:xi}) and
(\ref{Eq:hat-a})-(\ref{Eq:aa+1}) as follows.

By the same procedure leading to Eqs.~(\ref{Eq:aa+1}), we obtain
the general representation of the undeformed annihilation operator
\begin{equation}
\label{Eq:aa+2}
a_i=c_1^{\prime}(x_i
+ic_2^{\prime}p_i),\;c_1^{\prime}=\sqrt{\frac{1}{2\hbar
c_2^{\prime}}},
\end{equation}
The operators $a_i$ and $a_i^\dagger$ satisfy the undeformed
bosonic algebra
$$[a_{i},a_{j}]=[a_i^\dagger,a_j^\dagger]=0, \;
[a_{i},a^{\dagger}_{j}]=i\delta_{ij}.$$
The above undeformed bosonic commutation relation
$[a_{i},a_{j}]=0$ is automatically satisfied, so in
Eqs.~(\ref{Eq:aa+2}) the parameter $c_2^{\prime}$ is not
determined.

If NCQM is a realistic physics, the Hamiltonian of a system
reformulated in terms of deformed variables, like the Hamiltonian
(\ref{Eq:Hg}) of the quantum gravitational well in terms of
deformed variables $\hat x_i$ and $\hat p_i$, should have the same
representation as the one in terms of undeformed variables. At the
level of annihilation and creation operators in order to guarantee
that the Hamiltonian reformulated in terms of $\hat a_i$ and $\hat
a_i^{\dagger}$ has the same representation as the one formulated
in terms of $a_i$ and $a_i^{\dagger}$, the deformed $\hat a_i$ and
$\hat a_i^{\dagger}$ should have the same structure as the
undeformed $a_i$ and $a_i^{\dagger}$. Therefore, the parameters
$c_1^{\prime}$ and $c_2^{\prime}$ in Eqs.~(\ref{Eq:aa+2}) should
be the same ones $c_1$ and $c_2$ in Eqs.~(\ref{Eq:hat-a}):
\begin{equation}
\label{Eq:c1-c1}
c_1^{\prime}=c_1,\;c_2^{\prime}=c_2.
\end{equation}
Inserting Eqs.~(\ref{Eq:hat-x-p}) into Eqs.~(\ref{Eq:hat-a}), and
using Eqs.~(\ref{Eq:aa+2}) and (\ref{Eq:c1-c1}), we obtain
\begin{equation}
\label{Eq:hat-a-a1}
\hat a_{i}=\xi\left(a_{i}+\frac{i}{2\hbar}\sqrt{\theta\eta}
\epsilon_{ij}a_j\right),\;
\hat
a_{i}^\dagger=\xi\left(a_{i}^\dagger-\frac{i}{2\hbar}\sqrt{\theta\eta}
\epsilon_{ij}a_j^\dagger\right).
\end{equation}
All the deformed bosonic commutation relations in
(\ref{Eq:[a,a+]1}) are satisfied by Eqs.~(\ref{Eq:hat-a-a1});
Specially, $[\hat a_1,\hat a_1^\dagger]=[\hat a_2,\hat
a_2^\dagger]=1$ are maintained.

It is worth noting that the scaling factor $\xi$  guarantees
consistency of the framework, that is,
Eqs.~(\ref{Eq:xp})-(\ref{Eq:xi}),
(\ref{Eq:c-2})-(\ref{Eq:[a,a+]2}) and
(\ref{Eq:aa+2})-(\ref{Eq:hat-a-a1}) are consistent each other.

The determinant $\mathcal{R^{\; \prime}}_s$ of the transformation
matrix $R^{\; \prime}_s$ between $(\hat a_1,\hat a_2,\hat
a^\dagger_1,\hat a^\dagger_2)$ and
$(a_1,a_2,a^\dagger_1,a^\dagger_2)$ in Eqs.~(\ref{Eq:hat-a-a1}) is
also singular at $\theta\eta=4\hbar^2$. It means that $(\hat
a_i,\hat a^\dagger_i)$ and $(a_i,a^\dagger_i)$ are not completely
equivalent in physics.

{\it The tilde system} - In the tilde Fock space, as in
commutative space, all calculation is the same, thus the formalism
of the deformed Bosonic symmetry which restricts the states under
the permutation of identical particles in multi - particle systems
can be similarly developed.

Now we consider tilde phase space variables. Using
Eqs.~(\ref{Eq:aa+1}) and the definition (\ref{Eq:tilde-a}) of
$\tilde a_i$ we rewrite $\tilde a_i$ as
$\sqrt{\alpha_1}\; \tilde a_1 =\left
(\frac{\eta}{4\theta\hbar^2}\right)^{1/4}\left (\tilde x
+i\sqrt{\frac{\theta\hbar^2}{\eta}}\;\tilde
p^\dagger\right), \;$
$\sqrt{\alpha_2}\; \tilde a_2 =\left
(\frac{\eta}{4\theta\hbar^2}\right)^{1/4}\left (\tilde x^\dagger
+i\sqrt{\frac{\theta\hbar^2}{\eta}}\;\tilde p\right)$.
Where the tilde coordinate and momentum $(\tilde x, \tilde p)$ are
related to $(\hat x, \hat p)$ by
$\tilde x=\frac{1}{\sqrt{2}}\left (\hat x_1 +i\hat x_2\right)$,
$\tilde p=\frac{1}{\sqrt{2}}\left (\hat p_1-i\hat p_2\right)$.
The tilde phase variables $(\tilde x, \tilde p)$ satisfy the
following commutation relations:
$[\tilde x,\tilde x^\dagger]=\xi^2\theta, \;$
$[\tilde p,\tilde p^\dagger]= -\xi^2\eta, \;$
$[\tilde x,\tilde p]=[\tilde x^\dagger,\tilde p^\dagger]=i\hbar,
\;$
$[\tilde x,\tilde p^\dagger]=[\tilde x^\dagger,\tilde p]=0$.

The Hamiltonian $\hat H(\hat x,\hat p)=\frac{1}{2\mu}\hat p_i\hat
p_i +V(\hat x_i)$ with potential $V(\hat x_i)$ in the hat system
is rewritten as $\hat H(\hat x,\hat p)= \tilde H(\tilde x, \tilde
x^\dagger, \tilde p, \tilde p^\dagger) =\frac{1}{2\mu}\left
(\tilde p\tilde p^\dagger+ \tilde p^\dagger\tilde p\right)+\tilde
V(\tilde x, \tilde x^\dagger)$ in the tilde system. In some cases
calculations in the tilde system are simpler than ones in the hat
system.

Basis vectors of the tilde Fock space are the common eigen vectors
of commutative tilde number operators, so the tilde Fock space is
called as the commutative Fock space. Different from it,
Ref.~\cite{JLR} also investigated the structure of a
noncommutative Fock space, and obtained eigenvectors of several
pairs of commuting hermitian operators which can serve as basis
vectors in the noncommutative Fock space.

{\bf Constraint on Quantum Gravitational Well} - In the
constrained condition (\ref{Eq:cc1}) the direct proportional
coefficient $K$ is not determined. For the general case the
determination of $K$ or $c_2$ based on fundamental principles is
an open problem at present. Eqs.~(\ref{Eq:c1-c1}) represent the
condition that the Hamiltonian reformulated in terms of the
deformed annihilation and creation operators has the same
structure as the one formulated in terms of the undeformed
annihilation and creation operators. It shows that the
determination of $c_2$ or $K$ is related to the determination of
the coefficient $c_2^{\prime}$ of the undeformed annihilation
operator $a_{i}$.

Up to now the harmonic oscillator is the only example for which
the explicit representation of $c_2^{\prime}$ is known. At the
level of ordinary quantum mechanics the dimensional analysis works
for the determination of $c_2^{\prime}$ up to a dimensionless
constant. The dimension of $c_2^{\prime}$ in Eqs.~(\ref{Eq:aa+2})
is $time/mass$.

The special feature of a harmonic oscillator is that in any state
the expectation of the kinetic energy equals to the one of the
potential energy. This fixes the dimensionless constant $\gamma$.
The characteristic parameters in the Hamiltonian of a harmonic
oscillator are the mass $\mu$, frequency $\omega$ and $\hbar$. The
unique product of $\mu^{t_1}$, $\omega^{t_2}$ and $\hbar^{t_3}$
possessing the dimension $time/mass$ is $\mu^{-1}\omega^{-1}$. So
one obtains $c_2^{\prime}=\gamma/\mu\omega$. The position $x_{i}$
and momentum $p_{i}$ are, respectively, represented by $a_i$ and
$a_i^\dagger$ as
$$x_{i}=\sqrt{\frac{\gamma\hbar}{2\mu\omega}}\left(a_{i}+
a_i^\dagger\right),\;
p_{i}=-i\sqrt{\frac{\hbar\mu\omega}{2\gamma}}\left(a_{i}-
a_i^\dagger\right).$$
In the vacuum state $|0>$ the expectations of the kinetic and the
potential energy, respectively, read
$$\overline {E_k}=<0|\frac{1}{2\mu}p_{i}^2|0>
=\frac{\hbar\omega}{4\gamma},\;
\overline {E_p}=<0|\frac{1}{2}\mu\omega^2x_{i}^2|0>
=\frac{\gamma\hbar\omega}{4}.$$
The condition of $\overline {E_k}=\overline {E_p}$ leads to
$\gamma=\pm 1$. Because of $\overline {E_k}\ge 0$, the only
solution is $\gamma=1$.

The second example is a plain electromagnetic wave with a single
mode of frequency $\omega$. Its characteristic parameters are the
frequency $\omega$, the fundamental constant $\hbar$ and c (the
speed of light in vacuum). The constraint on it is
\begin{equation}
\label{Eq:cc3}
\eta=\kappa\hbar^2\omega^4 c^{-4} \theta.
\end{equation}
Here the coefficient $\kappa$ is an arbitrary dimensionless
constant.

Now we consider the quantum gravitational well.  Characteristic
parameters of the quantum gravitational well are the particle mass
$\mu_n$ and the standard gravitational acceleration $g$. Among
$\mu_n$, $g$, and the fundamental constants $\hbar$ and c (the
speed of light in vacuum) there are four possible combinations to
give the right dimension of $c_2^{\prime}$: $(\mu_n,g,\hbar)$,
$(\mu_n,g,c)$, $(\mu_n,\hbar,c)$ or $(g,\hbar,c)$. The deformed
Hamiltonian (\ref{Eq:Hg}) of the quantum gravitational well
includes parameters $\mu_n$, $g$ and $\hbar$. In order to obtain a
consistent representation of the corresponding Hamiltonian
formulated in terms of $a_i$ and $a_i^\dagger$, in the above we
should choose the first combination. This gives
$c_2^{\prime}=\gamma\left(\frac{\hbar}{\mu_n^4g^2}\right)^{1/3}.$
From Eqs.~(\ref{Eq:cc1}), (\ref{Eq:c1-c1}) and (\ref{Eq:c2}) it
follows that the constraint on the quantum gravitational well
reads
\begin{equation}
\label{Eq:cc2}
\eta=\zeta\left(\frac{\mu_n^4 g^2}{\hbar}\right)^{2/3}\theta,
\end{equation}
In the above the coefficient $\zeta$
is an arbitrary dimensionless constant.

The method of determining $\gamma$ for the harmonic oscillator can
not apply to the plain electromagnetic wave and the quantum
gravitational well.

\vspace{0.4cm}

We summarize the following points to conclude the paper.
(i) In the case of both position - position and momentum -
momentum noncommuting Bose - Einstein statistics is guaranteed by
the deformed Heisenberg - Weyl algebra itself, independent of
dynamics.
(ii) A special feature of a dynamical system is represented by a
constrained condition. A general feature of such a constraint for
any system is a direct proportionality between noncommutative
parameters with a coefficient depending on characteristic
parameters of the system under study. In the context of
non-relativistic quantum mechanics the dimensional analysis can
determine the proportional coefficient up to a dimensionless
constant. For the general case how to determine
such a dimensionless
constant is an open problem at present.
(iii) The discovery of the deformed correlated bosonic commutation
relation (\ref{Eq:[a,a+]2}) makes the deformed bosonic algebra
(\ref{Eq:[a,a+]1}) constituting a complete and closed algebra. In
literature one is not aware of this correlated bosonic commutation
relation. This means that in literature the deformed bosonic
algebra forms a closed algebra, but does not form a complete one.
(iv) The deformed annihilation and creation operators $\hat a_i$
and $\hat a_i^\dagger$ are represented by the undeformed ones
$a_i$ and $a_i^\dagger$ via the linear transformation
(\ref{Eq:hat-a-a1}), which maintains the bosonic commutation
relations, including $[\hat a_1,\hat a_1^\dagger]=[\hat a_2,\hat
a_2^\dagger]=1$.

\vspace{0.4cm}

The author would like to thank O. Bertolami for valuable
communications. This work has been supported by the Natural
Science Foundation of China under the grant number 10575037 and by
the Shanghai Education Development Foundation.



\begin{thebibliography}{99}
\bibitem{CDS}
A. Connes, M. R. Douglas, A. Schwarz,
JHEP {\bf 9802}, 003 (1998) {\bf hep-th/9711162}.

\bibitem{SW}
N. Seiberg and E. Witten,
JHEP {\bf 9909}:032 (1999) {\bf hep-th/9908142}.

\bibitem{DN}
M. R. Douglas, N. A. Nekrasov,
Rev. Mod. Phys. {\bf 73}, 977 (2001) {\bf hep-th/0106048} and
references there in.

\bibitem{CST}
M. Chaichian, M. M. Sheikh-Jabbari, A. Tureanu,
Phys. Rev. Lett. {\bf 86}, 2716 (2001) {\bf hep-th/0010175}.

\bibitem{GLR}
J. Gamboa, M. Loewe, J. C. Rojas,
Phys. Rev. {\bf D64}, 067901 (2001) {\bf hep-th/0010220}.

\bibitem{NP}
V. P. Nair, A. P. Polychronakos,
Phys. Lett. {\bf B505}, 267 (2001) {\bf hep-th/0011172}.

\bibitem{KD}
D. Kochan, M. Demetrian,
{\bf hep-th/0102050}

\bibitem{HK}
P-M. Ho, H-C. Kao,
Phys. Rev. Lett. {\bf 88}, 151602 (2002) {\bf hep-th/0110191}.

\bibitem{Bert05a}
O. Bertolami, J. G. Rosa, C. M. L. de Arag\~ao, P. Castorina,
D.Zappal\`a,
Phys. Rev. {\bf D72} 025010 (2005) {\bf hep-th/0505064}.

\bibitem{Bert05b}
O. Bertolami, J. G. Rosa, C. M. L. de Arag\~ao, P. Castorina,
D.Zappal\`a,
{\bf hep-th/0509207}.

\bibitem{JZZ04a}
Jian-zu Zhang,
Phys. Lett. {\bf B584}, 204 (2004) {\bf hep-th/0405135}.

\bibitem{JZZ04b}
Jian-zu Zhang,
Phys. Rev. Lett. {\bf 93}, 043002 (2004) {\bf hep-ph/0405143};

\bibitem{JZZ04c}
Jian-zu Zhang,
Phys. Lett. {\bf B597}, 362 (2004) {\bf hep-th/0407183}.

\bibitem{JZZ05a}
Qi-Jun Yin, Jian-zu Zhang,
Phys. Lett. {\bf B613}, 91 (2005) {\bf hep-th/0505246}.

\bibitem{JLR} 
S. C. Jin, Q. Y. Liu, T. N. Ruan,
{\bf hep-ph/0505048}.

\bibitem{Nesv02}
V. V. Nesvizhevsky et al., Nature {\bf 415} 297 (1995).

\bibitem{Nesv03}
V. V. Nesvizhevsky et al., Phys. Rev. {\bf D67} 102002 (2003).

\bibitem{Nesv05}
V. V. Nesvizhevsky et al., Eur. Phys. J, {\bf C40} 49 (2005) {\bf
hep-th/0502081}.

\bibitem{SGT}
T. P. Singh, S. Gutti, R. Tibrewala, {\bf gr-qc/0503116}.
\end{thebibliography}
\end{document}